\documentclass[
    ,final            
  ]
  {aipproc}

\layoutstyle{8x11single}
\usepackage{amssymb}
\usepackage{graphicx}
\usepackage{bm}

\begin{document}

\title{Transport Coefficients in Gluodynamics: From Weak Coupling
towards the Deconfinement Transition}


\author{M. Bluhm}{
  address={SUBATECH, UMR 6457, Universit\'{e} de Nantes,
Ecole des Mines de Nantes, IN2P3/CNRS. 4 rue Alfred Kastler,
F-44307 Nantes cedex 3, France}
}

\author{B. K\"ampfer}{
  address={Institut f\"ur Strahlenphysik,
Helmholtz-Zentrum Dresden-Rossendorf, PF 510119, D-01314 Dresden, Germany},
  altaddress={Institut f\"ur Theoretische Physik, TU Dresden,
D-01062 Dresden, Germany}
}

\author{K. Redlich}{
  address={Institute of Theoretical Physics, University of Wroclaw,
PL-50204 Wroclaw, Poland},
  altaddress={ExtreMe Matter Institute EMMI, GSI, D-64291 Darmstadt, Germany}
}

\begin{abstract}
We study the ratio of bulk to shear viscosity in gluodynamics within 
a phenomenological quasiparticle model. We show that at large temperatures 
this ratio exhibits a quadratic dependence on the conformality measure 
as known from weak coupling perturbative QCD. In the region of 
the deconfinement transition, however, this dependence becomes linear as 
known from specific strongly coupled theories. The onset of the 
strong coupling behavior is located near the maximum of the scaled 
interaction measure. This qualitative behavior of the viscosity ratio 
is rather insensitive to details of the equation of state. 
\end{abstract}

\maketitle


\section{Introduction}

Transport coefficients, such as bulk ($\zeta$) and shear ($\eta$) 
viscosities, represent fundamental quantities specifying the physical 
properties of the matter. In the hydrodynamic description 
of the medium created in high-energy nuclear collisions they are 
important input parameters. First-principle lattice QCD calculations 
of the viscosity coefficients have been performed 
recently in $SU_c(3)$ gluodynamics in the 
deconfinement region~\cite{Meyer,Sakai}. These lattice QCD results were 
shown to be successfully describable within a phenomenological 
quasiparticle model (QPM) for the gluon plasma~\cite{BluhmPRC,BluhmPLB}. 
This model is based on an effective kinetic theory approach to QCD and 
includes non-perturbative effects via a thermal quasigluon mass. In 
particular, in the vicinity of the deconfinement transition temperature 
$T_c$ the shear viscosity to entropy density ratio exhibits a 
minimum~\cite{BluhmPRC} with a value as small as the predicted universal 
lower bound~\cite{Kovtun} $\eta/s \sim  1/4\pi$. This result is in clear 
contrast to naive extrapolations of standard perturbative QCD 
approaches into the non-perturbative regime close to $T_c$. 
The specific bulk viscosity $\zeta/s$ is instead found to rapidly 
increase in the deconfinement transition region~\cite{BluhmPRC}. Moreover, 
the ratio $\zeta/\eta$ is known to exhibit a distinct dependence on the 
conformality measure $\Delta v_s^2=1/3-v_s^2$ with $v_s^2$ being 
the squared speed of sound. At weak coupling, i.e.~at high temperatures 
$T$, the viscosity ratio depends quadratically 
on $\Delta v_s^2$~\cite{Arnold}. On the other hand, at strong coupling, 
i.e.~close to $T_c$, this dependence is expected to be linear~\cite{hQCD}. 
The QPM reproduces the above asymptotic dependencies and exhibits a gradual 
interpolation between both regimes~\cite{BluhmPLB}. In this work, we analyze 
the structure of the viscosity ratio $\zeta/\eta$ in $SU_c(3)$ gluodynamics 
in the vicinity of $T_c$. In particular, we discuss its sensitivity on 
details in the equation of state (EoS) and the corresponding $v_s^2$. 

\section{Thermodynamics and Transport Coefficients}

In the QPM, the EoS, which describes changes of the pressure $P$ with energy 
density $\epsilon$, as well as related thermodynamic quantities are obtained 
from the local thermal equilibrium limit of the underlying effective kinetic 
theory. This model was shown to be very successful in describing the 
thermodynamic quantities obtained in lattice QCD~\cite{PRCPLBold}. 
Essential feature of the QPM is that the quasiparticle excitations 
follow a medium-modified dispersion relation, $E^2=\vec{p}^{\,2}+\Pi(T)$. In a 
quasigluon plasma, the effective mass $\Pi(T)=T^2G^2(T)/2$ is quantified by the 
coupling $G^2(T)=16\pi^2/\left(11\ln\left[\lambda(T-T_s)/T_c\right]^2\right)$, 
which near $T_c$ accommodates non-perturbative effects via only two parameters, 
$\lambda$ and $T_s$. At large $T$, it reproduces the perturbative running 
coupling in QCD. To determine $P$ and $\epsilon$ in the QPM, one needs to 
introduce the integration constant $B(T_c)$ as an additional 
parameter~\cite{PRCPLBold}. The squared speed of sound follows then from 
$v_s^2=\partial P/\partial\epsilon$. 

\begin{figure}[t]
  \includegraphics[height=.25\textheight]{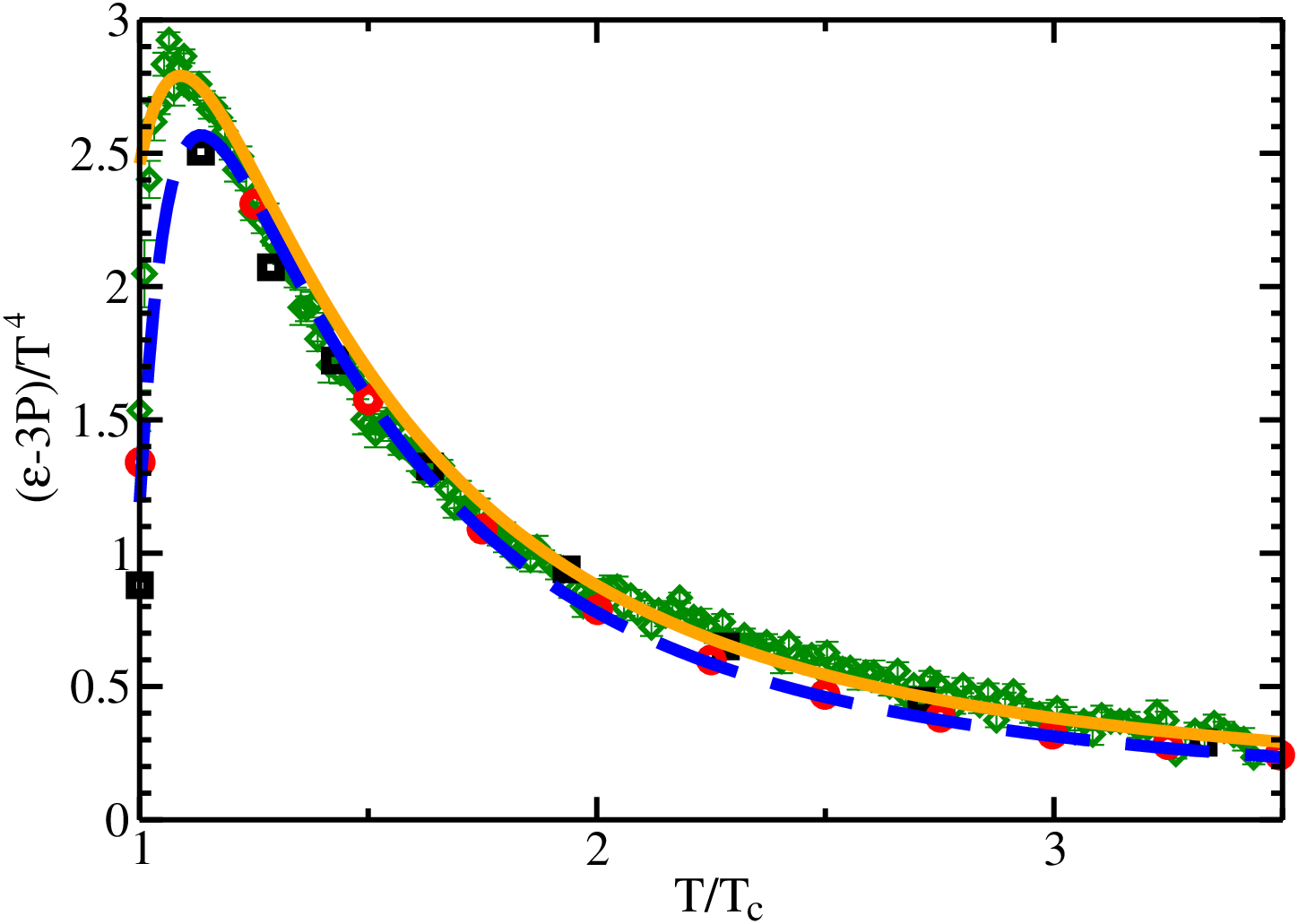}
  \hspace{2mm}
  \includegraphics[height=.25\textheight]{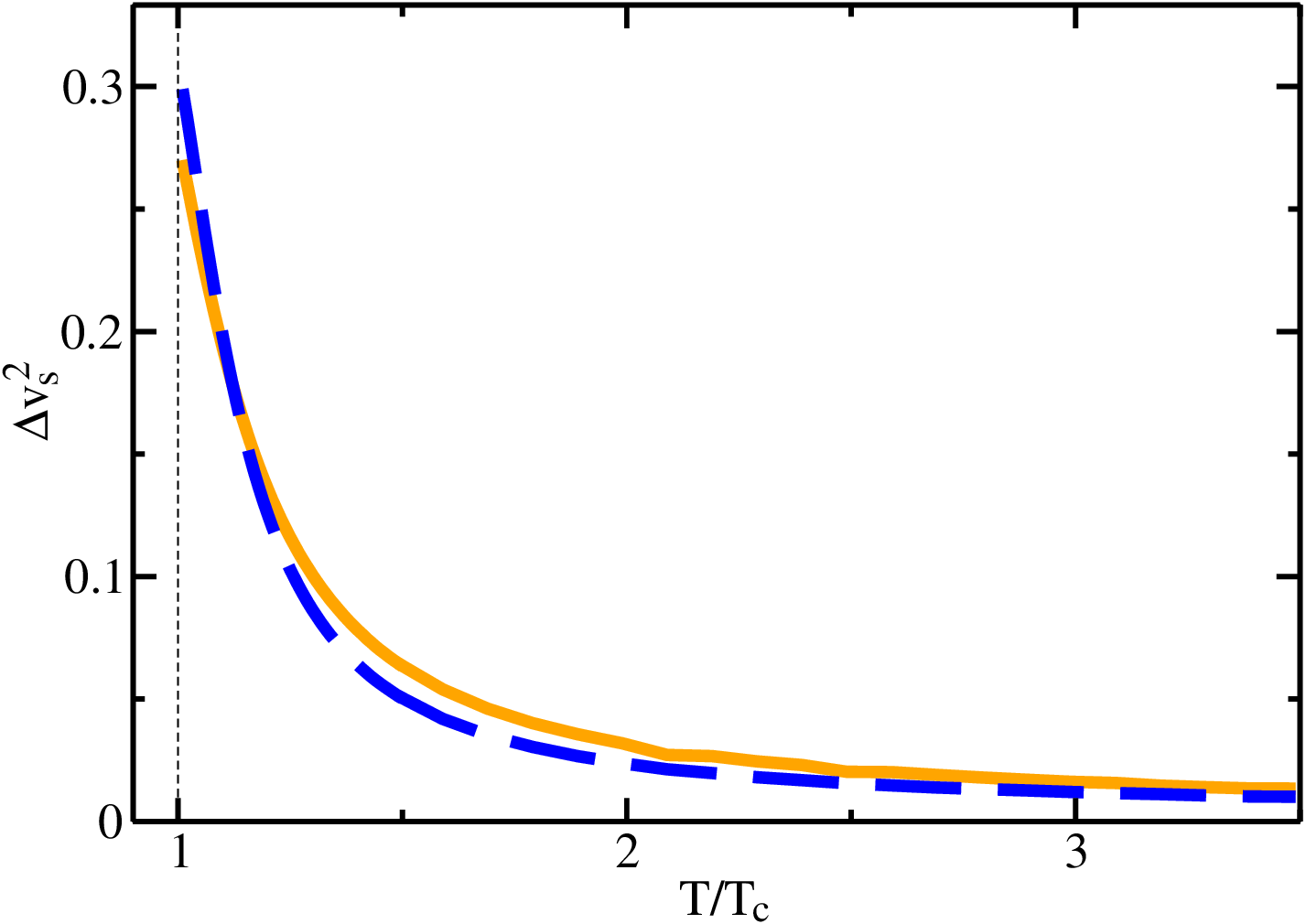}
  \caption{\label{fig:1} 
Left panel: Scaled interaction measure $(\epsilon-3P)/T^4$ as a function 
of the scaled temperature $T/T_c$ for $SU_c(3)$ gluodynamics. The 
symbols depict lattice QCD results from~\cite{Boyd} (squares),~\cite{Okamoto} 
(circles) and from~\cite{Panero} (diamonds). The dashed and solid 
curves are the QPM results for Fit~1 and Fit~2, respectively (see text). 
Right panel: Conformality measure $\Delta v_s^2$ as a function of $T/T_c$ in 
the QPM for Fit~1 (dashed curve) and Fit~2 (solid curve).}
\end{figure}
Figure~\ref{fig:1} (left panel) shows comparisons of the QPM predictions 
for the scaled interaction measure $(\epsilon-3P)/T^4$ as a function of 
$T/T_c$ with corresponding lattice QCD results~\cite{Boyd,Okamoto,Panero}. 
The model parameters read as $T_s/T_c=0.73$, $\lambda=4.3$ and 
$B(T_c)=0.19\,T_c^4$ (Fit 1) for data from~\cite{Boyd,Okamoto}, while we use 
$T_s/T_c=0.52$, $\lambda=2.5$ and $B(T_c)=0.48\,T_c^4$ (Fit 2) for data 
from~\cite{Panero}. The conformality measure obtained from these QPM results 
for the EoS is depicted in Fig.~\ref{fig:1} (right panel) as a function of $T/T_c$. 
The two different fits result in visible deviations in $\Delta v_s^2$ for 
$T_c<T\leq 2.5\,T_c$. 

At the first-order phase transition at $T=T_c$, the squared 
speed of sound $v_s^2(T)$ is discontinuous, which results in a discontinuity in 
$\Delta v_s^2$. We note that the limiting value 
$A=\lim_{T\to T_c^+} \Delta v_s^2$, which is linearly approached with 
$T/T_c$ as seen in Fig.~\ref{fig:1} (right panel), is in general different 
from $\Delta v_s^2(T_c)=1/3$. 

The bulk and shear viscosity coefficients follow directly from the effective 
kinetic theory~\cite{BluhmPRC,Kapusta}. Their ratio was found in the following 
form~\cite{BluhmPLB} 
\begin{eqnarray}
\label{equ:zetaetarationew}
 \frac{\zeta}{\eta} & = & 15\left(\Delta v_s^2\right)^2
 \left[1-\mathcal{A}_0+\frac14 \mathcal{A}_2 \right]
 + 5 \Delta v_s^2 \left[ \mathcal{A}_0- \frac12 \mathcal{A}_2 \right]
 + \frac{5}{12} \mathcal{A}_2 \,,
\end{eqnarray}
where 
$\mathcal{A}_0 = T^2\frac{dG^2}{dT^2} T^2\mathcal{I}_0/\mathcal{I}_{-2}$ and 
$\mathcal{A}_2 = \left(T^2\frac{dG^2}{dT^2}\right)^2 T^4\mathcal{I}_2/\mathcal{I}_{-2}$ 
depend non-trivially on $T$. 
The momentum integrals $\mathcal{I}_k$ in $\mathcal{A}_{0,2}$ read as 
$\mathcal{I}_k = \int \frac{d^3 \vec{p}}{(2\pi)^3}
n(T)[1+d^{-1}n(T)] \frac{\tau}{(E)^2}\vec{p}^{\,2-k}$, where 
$n(T)=d\left(\exp(E/T)-1\right)^{-1}$ is the Bose distribution 
function for gluons with $d=16$ and $\tau$ denotes the relaxation time. 
For $T\geq T_c$, these integrals follow the hierarchy 
$\mathcal{I}_{-2}/T^2\gg\mathcal{I}_0\gg T^2\mathcal{I}_2>0$, whereas 
$dG^2/dT^2<0$. Assuming a momentum independent $\tau$, the ratio $\zeta/\eta$ 
in Eq.~(\ref{equ:zetaetarationew}) is 
solely determined by parameters adjusted to equilibrium thermodynamics. 

\begin{figure}
  \includegraphics[height=.22\textheight]{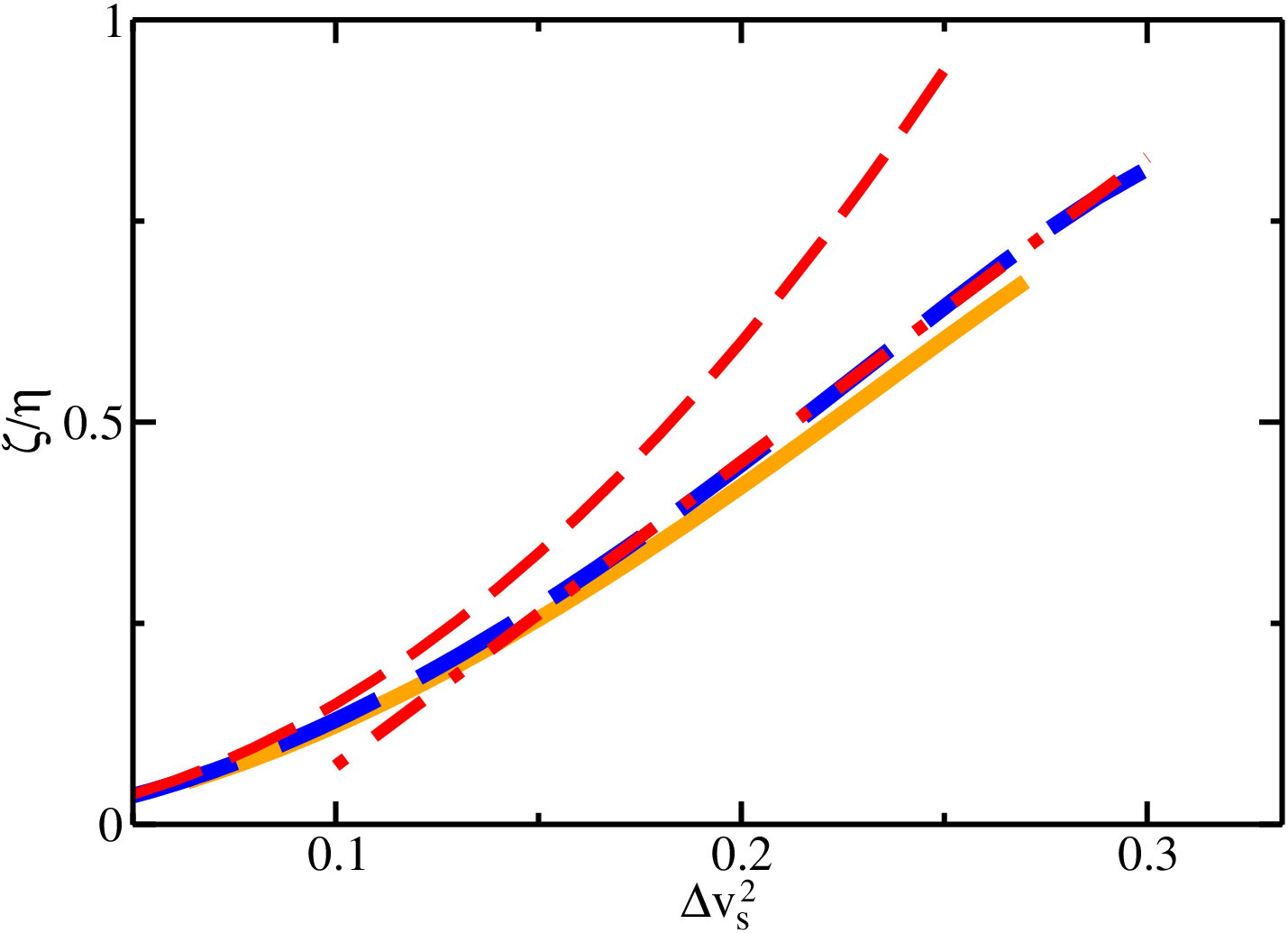}
  \caption{\label{fig:2} 
Bulk to shear viscosity ratio $\zeta/\eta$ from Eq.~(\ref{equ:zetaetarationew}) as a 
function of the conformality measure $\Delta v_s^2$ from Fig.~\ref{fig:1} (right panel) 
for Fit~1 (long-dashed curve) and Fit~2 (solid curve). The short-dashed curve exhibits 
the quadratic dependence $\zeta/\eta =15\left(\Delta v_s^2\right)^2$ and the dash-dotted 
curve shows a linear fit $\zeta/\eta=\alpha\,\Delta v_s^2+\beta$ of the QPM result from 
Fit~1 with $\alpha=3.78$ and $\beta=-0.305$. For the QPM result from Fit~2, a similar 
linear fit yields $\alpha=3.63$ and $\beta=-0.305$.}
\end{figure}
In Fig.~\ref{fig:2}, the ratio $\zeta/\eta$ from Eq.~(\ref{equ:zetaetarationew}) 
is quantified for Fit~1 and~2. For both QPM fits one observes that the viscosity 
ratio for $\Delta v_s^2 \lesssim 0.07$, i.e.~for $T\gtrsim 1.5\,T_c$, is entirely 
determined by $15\left(\Delta v_s^2\right)^2$, i.e.~by the quadratic, 
$\mathcal{A}_{0,2}$-independent term in Eq.~(\ref{equ:zetaetarationew}). This is a 
direct consequence of the behavior of the conformality measure at large $T$, which 
in the QPM reads as $\Delta v_s^2\simeq -75/(18\,d\pi^2)T^2(dG^2/dT^2)+
\mathcal{O}(G^2T^2(dG^2/dT^2))$ with $\vert T^2(dG^2/dT^2)\vert\ll G^2\ll 1$. 
Thus, at leading order, all terms in Eq.~(\ref{equ:zetaetarationew}) 
depend quadratically on $\Delta v_s^2$, where the first, $\mathcal{A}_{0,2}$-independent 
term dominates numerically. With increasing $\Delta v_s^2$, the 
$\zeta/\eta$ ratio is reduced compared to $15\left(\Delta v_s^2\right)^2$ by the 
non-perturbative terms in Eq.~(\ref{equ:zetaetarationew}) which are proportional to 
$(dG^2/dT^2)$. Consequently, one finds $\zeta/\eta<1$ for 
$T\to T_c^+$. For $\Delta v_s^2\gtrsim 0.17$, i.e.~for $T\lesssim 1.15\,T_c$, a linear 
dependence on $\Delta v_s^2$ develops as seen in Fig.~\ref{fig:2}. In fact, near $T_c$ 
the factors $\vert\mathcal{A}_{0,2}\vert$ become large as a consequence of a large 
$\vert dG^2/dT^2\vert$. This results in a cancelation of the quadratic dependence 
$15\left(\Delta v_s^2\right)^2$ in Eq.~(\ref{equ:zetaetarationew}) by all 
other $\mathcal{A}_{0,2}$-dependent terms. Effectively, Eq.~(\ref{equ:zetaetarationew}) 
sums up near $T_c$ to a linear dependence of the form 
$\zeta/\eta=\alpha\,\Delta v_s^2 +\beta$. The transition to this linear behavior, as it 
is known for specific strongly coupled theories~\cite{hQCD}, takes place at 
$T$ in the vicinity of the maximum in the scaled interaction measure, cf.~Fig.~\ref{fig:1}. 
As is evident from Fig.~\ref{fig:2}, this qualitative behavior is a common feature 
of the viscosity ratio within the quasiparticle model for the gluon plasma irrespective of 
details in the EoS near $T_c$. In the interval $0.07\lesssim\Delta v_s^2\lesssim 0.17$, 
a gradual change between quadratic and linear dependence on $\Delta v_s^2$ 
is observed. We note that at $T=T_c$ the $\zeta/\eta$ ratio develops a discontinuity 
in line with $\Delta v_s^2$. 

\section{Summary}

Within a phenomenological quasiparticle model for the gluon plasma, the 
bulk to shear viscosity ratio is found to exhibit, at large $T$, the quadratic 
dependence on the conformality measure as known from perturbative QCD. In the 
deconfinement transition region, this dependence becomes linear as found 
in specific strongly coupled theories. Thus, the quasiparticle model provides 
a systematic link between both regimes. The onset of the strong coupling behavior 
is located near the maximum in the scaled interaction measure. This qualitative 
behavior of the viscosity ratio is insensitive to details in the equation of state. 

\noindent
\textbf{Acknowledgments.}
We kindly acknowledge stimulating discussions with A.~Buchel, S.~Jeon, 
H.~B.~Meyer, J.~Noronha, S.~Peign\'{e} and A.~Peshier. We also thank M.~Panero 
for providing his lattice QCD results. M. Bluhm thanks the organizers of 
the PANIC11 conference for financial support. 
This work is supported by BMBF 06DR9059, GSI-FE, the European Network 
I3-HP2 Toric and the Polish Ministry of Science. 





\bibliographystyle{aipproc}   



\end{document}